# A closed-form GN-model Non-Linear Interference Coherence Term

*Pierluigi Poggiolini*
*Senior Member, IEEE*
*Fellow, Optical Society of America (OSA)*
Politecnico di Torino, Italy
pierluigi.poggiolini@polito.it

**Abstract:** In this paper we report the details of the derivation of approximate closed-form results related to the coherence effect in non-linear noise accumulation, in the context of the GN-model of fiber non-linearity. The coherence effect is particularly important in relation to the non-linearity noise produced by a single transmission channel onto itself. We derive new results, not shown before, that provide a more accurate representation of coherent accumulation of NLI than previous closed-form formulas. The availability of such closed-form formulas is important in the context of real-time management and optimization of physical-layer aware optical networks.

## 1. Introduction

The problem of modeling the impact of non-linear effects in an optical fiber has been paid substantial attention since the onset of optical fiber communications. Currently, several NLI (non-linear-interference) models are available, with different features in terms of accuracy vs. complexity, see for instance [1]-[8].

One of the most well-known and one of the most widely used in the industry is the so-called "GN-model". Interestingly, while its name and its current evolved form(s) are relatively recent, the first derivation of an NLI model along similar lines as the GN-model dates back to 1994 [9]. For a comprehensive introduction to the GN-model, its history, evolution, validation and limitations, please see [5],[10].

Based on the GN model, various closed-form formulas expressing NLI were derived, for instance in [11]-[14]. Lately, the effort towards finding more general and effective closed-form formulas has been revamped and several papers have been published on this topic, among which [15]-[19]. The need for new closed-form formulas is generated by the push by the industry and operators to have at their disposal real-time accurate tools to support physical layer-aware optical network management and real-time optimization.

This paper provides results that are meant to complement those of [11] and [16],[17],[20]. Specifically, we first provide a detailed re-derivation of some results presented in [11]. These results deal with *improving accuracy* of closed-form GN-model formulas by taking into account the so-called *coherent accumulation* of NLI.

Then we specialize to the case of single-channel non-linearity, or single-channel interference (SCI), whereby we concentrate on the non-linearity produced by a single channel onto itself. In this context we propose new unpublished results.

We then link such results to the effort carried out in [16],[17],[20], aimed at obtaining an all-encompassing accurate closed-form GN-model formula capable of dealing with a very wide variety of practical systems. This last passage required substantial approximations and drastic assumptions. However, the findings in [20], where the *approximate coherent accumulation correction term* derived here is used, appear to confirm its beneficial effect towards improving accuracy.

This document is an update to the previous version. The update consists of a typo being corrected in Eq.(39) of the previous version (now Eq.(40)), which affected other formulas downstream as well.

## 2. Closed-form GN-model formula for the Nyquist-WDM case

In this section we derive an approximate closed-form *coherence correction term* to be used to improve the accuracy of the GN-model closed-form formulas that neglect the *coherent accumulation of NLI*. The latter consists of the coherent beating of NLI contributions produced in each span of a link, occurring at the end of the link, where the receiver is located. If this effect is neglected, the *incoherent* GN-model is obtained, which assumes that the NLI contributions of each span in the link simply sum up *in power* at the receiver. Notice that the integral GN-model Reference Formula (GNRF, see Eqs. 1-3 in [11]) does take coherent NLI accumulation into account. However, integral formulas are cumbersome to use as they require lengthy numerical integration. The general goal is to achieve closed-form expressions that allow real-time use in practical networking applications.

Notice that the derivation shown in the following leads to Eqs. 22-24 in [11] starting from Eq. 21 in [11]. It was proposed in concise form as Appendix H of [11]. It is re-proposed here in significantly extended form, to facilitate the subsequent derivation of new unpublished results in the next sections.

We start out from the GRNF Eqs. 1-3 in [11]. We make the underlying assumption that the spans in the link are all identical, with lumped amplification. The optical transmission Power Spectral Density (PSD) is $G_{\text{WDM}}(f)$ and it is assumed to be of ideal rectangular shape and centered at (offset) frequency $f = 0$, that is:

$$G_{\text{WDM}}(f) = G_0 \cdot \Pi_{B_{\text{WDM}}}(f)$$

*Eq. 1*

where $G_0$ is a constant and the function $\Pi_{B_{\text{WDM}}}(f)$ is a Heaviside Pi function, defined here as being 0 for $f \notin [-B_{\text{WDM}}/2, B_{\text{WDM}}/2]$ and 1 otherwise. This assumption on $G_{\text{WDM}}(f)$ may correspond to: (i) an ideal Nyquist-WDM multi-channel spectrum, with channel spacing equal to the symbol rate and either offset-transmission or zero roll-off factor; (ii) a single optical channel with rectangular spectrum. We will later provide comments regarding this distinction, which for now does not need to be made.

We want to characterize the PSD of the Non-Linear Interference (NLI) noise at the center of the optical transmission spectrum $f = 0$, i.e, we want to calculate $G_{\text{NLI}}(0)$. Note that characterizing NLI at the center frequency may be enough to obtain an approximate overall spectrum characterization, under the assumption of Eq. 1 and the *locally white NLI noise approximation* (see [11]) consisting of considering NLI noise spectrally flat over the bandwidth of the Nyquist-WDM (or single-channel) signal.

Combining the GNRF Eqs. 1-3 from [11] with the assumption on the signal PSD Eq. 1, we can rewrite the GNRF as:

$$G_{\text{NLI}}(f) = \frac{16}{27} \gamma^2 \int_{-\infty}^{\infty} \int_{-\infty}^{\infty} \left| \frac{1 - e^{-2\alpha L_s} e^{j4\pi^2 \beta_2 L_s (f_1 - f)(f_2 - f)}}{2\alpha - j4\pi^2 \beta_2 (f_1 - f)(f_2 - f)} \right|^2 \cdot \frac{\sin^2\left(2N_s \pi^2 (f_1 - f)(f_2 - f) \beta_2 L_s\right)}{\sin^2\left(2\pi^2 (f_1 - f)(f_2 - f) \beta_2 L_s\right)} \cdot$$
$$\cdot G_{\text{WDM}}(f_1) G_{\text{WDM}}(f_2) G_{\text{WDM}}(f_1 + f_2 - f) \, df_1 df_2$$

*Eq. 2*

The notation used here is the same as in [11]. The symbols have the following meaning:

- $L_s$ : span length
- $2\alpha$ : fiber power loss coefficient (km$^{-1}$), such that signal power is attenuated over a span as $\exp(-2\alpha \cdot L_s)$
- $\beta_2$ : absolute value (always positive) of fiber dispersion (ps$^2$)
- $\gamma$ : fiber Kerr non-linearity coefficient (W·km)$^{-1}$
- $L_{\text{eff}}$ : non-linearity effective length (km), defined as $L_{\text{eff}} = [1 - \exp(-2\alpha \cdot L_s)]/(2\alpha)$
- $N_s$ : number of spans in the link

The factor in the integral within absolute value represents the Four-Wave Mixing (FWM) power efficiency of the beating of signal optical power present at the three frequencies $f_1$, $f_2$, and $(f_1 + f_2 - f)$, creating NLI power at frequency $f$. The factor following the FWM factor is a form of the Fejér kernel [21], that has interesting properties that we will use later.

Substituting Eq. 1 into Eq. 2 and imposing $f = 0$, we get:

$$G_{\text{NLI}}(0) = \frac{16}{27}\gamma^2 G_0^3 \int_{-\infty}^{\infty}\int_{-\infty}^{\infty} \left|\frac{1-e^{-2\alpha L_s}e^{j4\pi^2\beta_2 L_s f_1 f_2}}{2\alpha - j4\pi^2\beta_2 f_1 f_2}\right|^2 \cdot \frac{\sin^2(2N_s\pi^2 f_1 f_2 \beta_2 L_s)}{\sin^2(2\pi^2 f_1 f_2 \beta_2 L_s)} \cdot$$

$$\cdot \Pi_{B_{\text{WDM}}}(f_1)\Pi_{B_{\text{WDM}}}(f_2)\Pi_{B_{\text{WDM}}}(f_1+f_2)\,df_1\,df_2$$

*Eq. 3*

The factors $\Pi_{B_{\text{WDM}}}(f_1)$ and $\Pi_{B_{\text{WDM}}}(f_2)$ are zero for $f_1 \notin [-B_{\text{WDM}}/2, B_{\text{WDM}}/2]$ and $f_2 \notin [-B_{\text{WDM}}/2, B_{\text{WDM}}/2]$, respectively, and 1 otherwise. As a result, they curtail the integration ranges and disappear from the integrand function, as follows:

$$G_{\text{NLI}}(0) = \frac{16}{27}\gamma^2 G_0^3 \int_{-B_{\text{WDM}}/2}^{B_{\text{WDM}}/2}\int_{-B_{\text{WDM}}/2}^{B_{\text{WDM}}/2} \left|\frac{1-e^{-2\alpha L_s}e^{j4\pi^2\beta_2 L_s f_1 f_2}}{2\alpha - j4\pi^2\beta_2 f_1 f_2}\right|^2 \cdot \frac{\sin^2(2N_s\pi^2 f_1 f_2 \beta_2 L_s)}{\sin^2(2\pi^2 f_1 f_2 \beta_2 L_s)} \cdot \Pi_{B_{\text{WDM}}}(f_1+f_2)\,df_1\,df_2$$

*Eq. 4*

The FWM efficiency factor can be rewritten as:

$$\left|\frac{1-e^{-2\alpha L_s}e^{j4\pi^2\beta_2 L_s f_1 f_2}}{2\alpha - j4\pi^2\beta_2 f_1 f_2}\right|^2 = \frac{\left(1-e^{-2\alpha L_s}\right)^2 + 4e^{-2\alpha L_s}\sin^2(2\pi^2\beta_2 f_1 f_2 L_s)}{4\alpha^2 + 16\pi^4\beta_2^2 f_1^2 f_2^2}$$

*Eq. 5*

Provided that the span loss is large enough, indicatively about 10 dB or greater, then we can perform the following approximation:

$$\left|\frac{1-e^{-2\alpha L_s}e^{j4\pi^2\beta_2 L_s f_1 f_2}}{2\alpha - j4\pi^2\beta_2 f_1 f_2}\right|^2 = \frac{\left(1-e^{-2\alpha L_s}\right)^2 + 4e^{-2\alpha L_s}\sin^2(2\pi^2\beta_2 f_1 f_2 L_s)}{4\alpha^2 + 16\pi^4\beta_2^2 f_1^2 f_2^2} \approx$$

$$\approx \frac{\left(1-e^{-2\alpha L_s}\right)^2}{4\alpha^2 + 16\pi^4\beta_2^2 f_1^2 f_2^2} = \frac{L_{\text{eff}}^2}{1+16\pi^4(2\alpha)^{-2}\beta_2^2 f_1^2 f_2^2}$$

*Eq. 6*

This leads to:

$$G_{\text{NLI}}(0) \approx \frac{16}{27}\gamma^2 L_{\text{eff}}^2 G_0^3 \int_{-B_{\text{WDM}}/2}^{B_{\text{WDM}}/2}\int_{-B_{\text{WDM}}/2}^{B_{\text{WDM}}/2} \frac{1}{1+16\pi^4(2\alpha)^{-2}\beta_2^2 f_1^2 f_2^2}$$

$$\cdot \frac{\sin^2(2N_s\pi^2 f_1 f_2 \beta_2 L_s)}{\sin^2(2\pi^2 f_1 f_2 \beta_2 L_s)} \cdot \Pi_{B_{\text{WDM}}}(f_1+f_2)\,df_1\,df_2$$

*Eq. 7*

The integration domain resulting from the integration limits shown above and the presence in the integrand function of $\Pi_{B_{\text{WDM}}}(f_1+f_2)$, which is 0 for $(f_1+f_1) \notin [-B_{\text{WDM}}/2, B_{\text{WDM}}/2]$, results in lozenge-shaped region over the $(f_1, f_1)$ plane, as shown in Fig. 1.

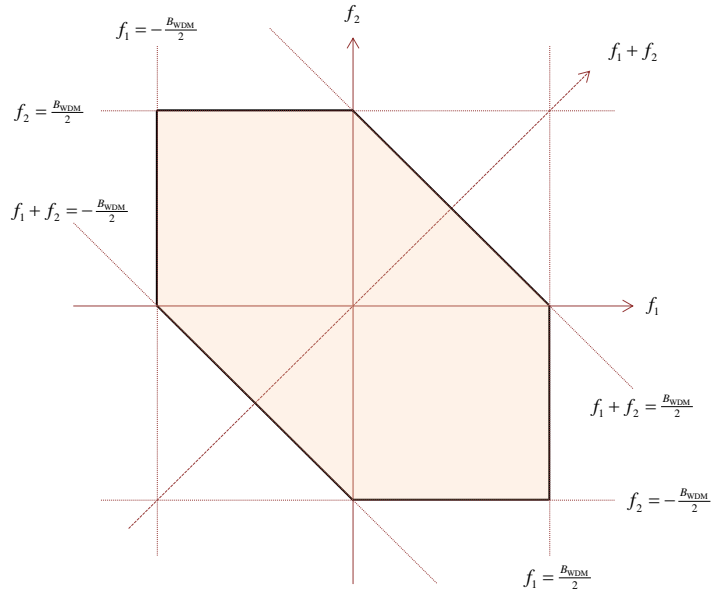

*Fig. 1 - shaded area: lozenge-shaped exact integration domain of Eq. 7)*

Taking the exact integration domain into account would however complicate the calculations substantially. As a result, we perform an approximation, consisting of assuming a perfectly square integration domain, as shown in Fig. 2. This is equivalent to neglecting in the integrand the factor $\Pi_{B_{\text{WDM}}}(f_1 + f_2)$, which is the one generating the two slanted 45° borders.

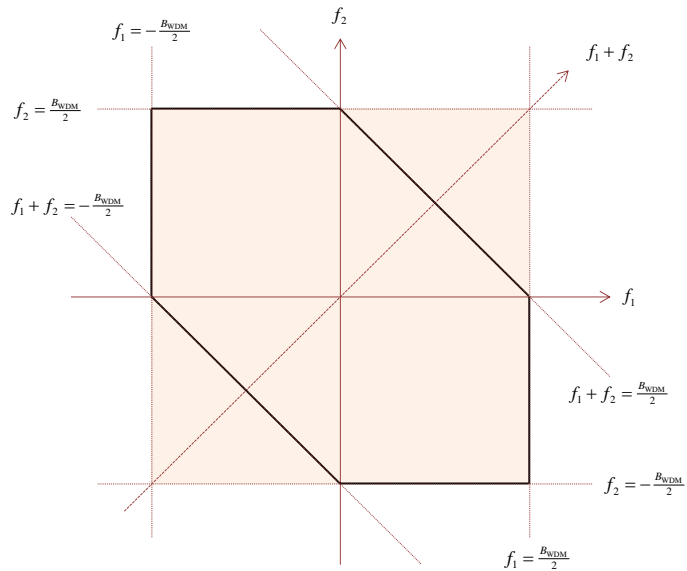

*Fig. 2 – shaded area: approximate square integration domain*

This approximation is justified by various considerations. First, since the integrand is everywhere positive, then increasing the area of integration leads to overestimating NLI, i.e., to conservative (pessimistic) system performance predictions. In addition, it is easy to see that such overestimation is bounded by a factor 4/3 (the ratio of the square vs. the lozenge). The value 4/3 is however reached only if $\beta_2 = 0$.

Other choices of integrand region manipulation are possible. In [11], Appendix F, a circular approximation to the lozenge of Fig. 1 is performed. Here we adopt the rectangular approximation of Fig. 2.

Analytically, performing such approximation is equivalent to removing $\Pi_{B_{\text{WDM}}}(f_1 + f_2)$ from the integrand. We can then write:

$$G_{\text{NLI}}(0) = \frac{16}{27} \gamma^2 L_{\text{eff}}^2 G_0^3 \int_{-B_{\text{WDM}}/2}^{B_{\text{WDM}}/2} \int_{-B_{\text{WDM}}/2}^{B_{\text{WDM}}/2} \frac{1}{1+16\pi^4(2\alpha)^{-2}\beta_2^2 f_1^2 f_2^2} \cdot \frac{\sin^2\left(2N_s\pi^2 f_1 f_2 \beta_2 L_s\right)}{\sin^2\left(2\pi^2 f_1 f_2 \beta_2 L_s\right)} df_1 df_2$$

*Eq. 8*

We then elect to carry out the first integration over the variable $f_1$. The integral on $f_1$ is:

$$Q_1 = \int_{-B_{\text{WDM}}/2}^{B_{\text{WDM}}/2} \frac{1}{1+16\pi^4(2\alpha)^{-2}\beta_2^2 f_1^2 f_2^2} \cdot \frac{\sin^2\left(2N_s\pi^2 f_1 f_2 \beta_2 L_s\right)}{\sin^2\left(2\pi^2 f_1 f_2 \beta_2 L_s\right)} df_1$$

*Eq. 9*

For convenience we define the asymptotic effective length:

$$L_\infty = \lim_{L_s \to \infty} L_{\text{eff}} = \lim_{L_s \to \infty} \frac{1-\exp(-2\alpha \cdot L_s)}{2\alpha} = \frac{1}{2\alpha}$$

*Eq. 10*

This would be the effective length of an infinitely long fiber span. We substitute and write:

$$Q_1 = \int_{-B_{\text{WDM}}/2}^{B_{\text{WDM}}/2} \frac{1}{1+16\pi^4 \beta_2^2 L_\infty^2 f_1^2 f_2^2} \cdot \frac{\sin^2\left(2N_s\pi^2 f_1 f_2 \beta_2 L_s\right)}{\sin^2\left(2\pi^2 f_1 f_2 \beta_2 L_s\right)} df_1$$

We also define for notational convenience:

$$h = 2\pi^2 f_2 \beta_2 L_s$$

*Eq. 11*

from which:

$$Q_1 = \int_{-B_{\text{WDM}}/2}^{B_{\text{WDM}}/2} \frac{1}{1+4\cdot 4\pi^4 \beta_2^2 f_2^2 L_s^2 \frac{L_\infty^2}{L_s^2} f_1^2} \cdot \frac{\sin^2(hN_s f_1)}{\sin^2(h f_1)} df_1 = \int_{-B_{\text{WDM}}/2}^{B_{\text{WDM}}/2} \frac{1}{1+4\cdot h^2 \frac{L_\infty^2}{L_s^2} f_1^2} \cdot \frac{\sin^2(hN_s f_1)}{\sin^2(h f_1)} df_1$$

$$= \int_{-B_{\text{WDM}}/2}^{B_{\text{WDM}}/2} \frac{\frac{L_s^2}{4L_\infty^2}}{\frac{L_s^2}{4L_\infty^2} 1+4\cdot h^2 \frac{L_\infty^2}{L_s^2} f_1^2} \cdot \frac{\sin^2(hN_s f_1)}{\sin^2(h f_1)} df_1 = \frac{L_s^2}{4L_\infty^2} \int_{-B_{\text{WDM}}/2}^{B_{\text{WDM}}/2} \frac{1}{\frac{L_s^2}{4L_\infty^2}+h^2 f_1^2} \cdot \frac{\sin^2(hN_s f_1)}{\sin^2(h f_1)} df_1$$

*Eq. 12*

Again, defining for notational convenience:

$$b^2 = \frac{L_s^2}{4L_\infty^2} \quad , \quad q = \frac{B_{\text{WDM}}}{2}$$

*Eq. 13*

we get:

$$Q_1 = \int_{-q}^{q} \frac{b^2}{b^2 + h^2 f_1^2} \cdot \frac{\sin^2(hN_s f_1)}{\sin^2(h f_1)} df_1$$

*Eq. 14*

One interesting property of the Fejér kernel factor is that it can be rewritten in the form of a finite series as follows:

$$\frac{\sin^2(Nx)}{\sin^2(x)} = N + 2\sum_{n=1}^{N-1}(N-n)\cdot\cos(2n\cdot x)$$

<div align="center">Eq. 15</div>

So, we can re-write:

$$Q_1 = \int_{-q}^{q} \frac{b^2}{b^2+h^2 f_1^2} \cdot \frac{\sin^2(hN_s f_1)}{\sin^2(hf_1)} df_1 =$$

$$= \int_{-q}^{q} \frac{b^2}{b^2+h^2 f_1^2} \cdot \left[N_s + 2\sum_{n=1}^{N_s-1}(N_s-n)\cdot\cos(2n\cdot hf_1)\right] df_1 =$$

$$= N_s \int_{-q}^{q} \frac{b^2}{b^2+h^2 f_1^2} df_1 + 2\sum_{n=1}^{N_s-1}(N_s-n)\int_{-q}^{q} \frac{b^2 \cos(2n\cdot hf_1)}{b^2+h^2 f_1^2} df_1$$

<div align="center">Eq. 16</div>

Eq. 16) contains two integrals. The first integral can be solved resorting to the general formula:

$$\int \frac{1}{1+\rho^2 x^2} dx = \frac{\tan^{-1}(\rho x)}{\rho}$$

<div align="center">Eq. 17</div>

from which:

$$N_s \int_{-q}^{q} \frac{b^2}{b^2+h^2 f_1^2} df_1 = N_s \int_{-q}^{q} \frac{1}{1+\frac{h^2}{b^2} f_1^2} df_1 = N_s \left[\frac{\tan^{-1}\left(\frac{h}{b} f_1\right)}{\frac{h}{b}}\right]_{-q}^{q} = \frac{2bN_s}{h}\tan^{-1}\left(\frac{hq}{b}\right)$$

<div align="center">Eq. 18</div>

Note that the sign of $h$ in the result is irrelevant, because the integrand function is even in $h$. To stress this circumstance we re-write the result as:

$$N_s \int_{-q}^{q} \frac{b^2}{b^2+h^2 f_1^2} df_1 = \frac{2bN_s}{|h|}\tan^{-1}\left(\frac{|h|q}{b}\right)$$

<div align="center">Eq. 19</div>

The second integral in Eq. 16 also has an analytical solution:

$$\int_{-q}^{q} \frac{b^2 \cos(2n\cdot hf_1)}{b^2+h^2 f_1^2} df =$$

$$= \frac{1}{h} jb\cosh(2bn)\{\text{Ci}(2n[jb+hq]) - \text{Ci}(2n[-jb+hq])\} +$$

$$+ \frac{\pi}{h} b\cosh(2bn) + \frac{1}{h} b\sinh(2bn)\{\text{Si}(2n[jb-hq]) - \text{Si}(2n[jb+hq])\}$$

$$= \frac{b}{h}\{\cosh(2bn)[j\text{Ci}(2n[jb+hq]) - j\text{Ci}(2n[-jb+hq]) + \pi] + \sinh(2bn)[\text{Si}(2n[jb-hq]) - \text{Si}(2n[jb+hq])]\}$$

<div align="center">Eq. 20</div>

The above formula is valid for all real values of $h$, both positive and negative.

An equivalent and more compact form can be obtained in terms of Exponential-Integral functions:

$$\int_{-q}^{q} \frac{b^2 \cos(2n \cdot hf_1)}{b^2 + h^2 f_1^2} df = \frac{b}{2h} e^{-2nb} \left[ -je^{4nb} \text{Ei}(-2n[b+jhq]) - j\text{Ei}(2n[b+jhq]) \right.$$
$$\left. + je^{4nb} \text{Ei}(-2n[b-jhq]) + j\text{Ei}(2n[b-jhq]) + 2\pi e^{4nb} \right] \quad , \quad h > 0$$

*Eq. 21*

This expression is valid only for $h > 0$. On the other hand, looking at the integrand function in:

$$\int_{-q}^{q} \frac{b^2 \cos(2n \cdot hf_1)}{b^2 + h^2 f_1^2} df$$

it is apparent that the sign of $h$ is irrelevant, because the integrand function is even vs. $h$. Therefore, we can write:

$$\int_{-q}^{q} \frac{b^2 \cos(2n \cdot hf_1)}{b^2 + h^2 f_1^2} df = \frac{b}{2|h|} e^{-2nb} \left[ -je^{4nb} \text{Ei}(-2n[b+j|h|q]) - j\text{Ei}(2n[b+j|h|q]) \right.$$
$$\left. + je^{4nb} \text{Ei}(-2n[b-j|h|q]) + j\text{Ei}(2n[b-j|h|q]) + 2\pi e^{4nb} \right]$$

*Eq. 22*

We then remark that: $\text{Ei}(z) = \text{Ei}^*(z^*)$, and exploiting this identity we can rewrite Eq. 22) as:

$$\int_{-q}^{q} \frac{b^2 \cos(2n \cdot hf_1)}{b^2 + h^2 f_1^2} df = \frac{b}{|h|} \left[ -e^{2nb} \text{Im}\{\text{Ei}(-2n[b-j|h|q])\} - \right.$$
$$\left. e^{-2nb} \text{Im}\{\text{Ei}(2n[b-j|h|q])\} + \pi e^{2nb} \right]$$

*Eq. 23*

Therefore, Eq. 16 can be fully analytically integrated, yielding:

$$Q_1 = \int_{-q}^{q} \frac{b^2}{b^2 + h^2 f_1^2} \cdot \frac{\sin^2(hN_s f_1)}{\sin^2(hf_1)} df_1 = \frac{2bN_s}{|h|} \tan^{-1}\left(\frac{|h|q}{b}\right) +$$
$$2\frac{b}{|h|} \sum_{n=1}^{N_s-1} (N_s - n) \left[ -e^{2nb} \text{Im}\{\text{Ei}(-2n[b-j|h|q])\} - e^{-2nb} \text{Im}\{\text{Ei}(2n[b-j|h|q])\} + \pi e^{2nb} \right]$$

*Eq. 24*

We now recall the substitutions that we made for notational convenience:

$$b^2 = \frac{L_s^2}{4L_\infty^2} \quad q = \frac{B_{\text{WDM}}}{2} \quad h = 2\pi^2 f_2 \beta_2 L_s \quad h^2 = 4\pi^4 f_2^2 \beta_2^2 L_s^2 \quad |h| = 2\pi^2 |f_2| \beta_2 L_s$$

and undo them in Eq. 24, obtaining:

$$Q_1 = \frac{2\frac{L_s}{2L_\infty}N_s}{2\pi^2|f_2|\beta_2 L_s}\tan^{-1}\left(\frac{2\pi^2|f_2|\beta_2 L_s\frac{B_{\text{WDM}}}{2}}{\frac{L_s}{2L_\infty}}\right) + 2\frac{\frac{L_s}{2L_\infty}}{2\pi^2|f_2|\beta_2 L_s}\sum_{n=1}^{N_s-1}(N_s-n)$$

$$\left[-e^{2n\frac{L_s}{2L_\infty}}\text{Im}\left\{\text{Ei}\left(-2n\left[\frac{L_s}{2L_\infty}-j2\pi^2|f_2|\beta_2 L_s\frac{B_{\text{WDM}}}{2}\right]\right)\right\}-\right.$$

$$\left.e^{-2n\frac{L_s}{2L_\infty}}\text{Im}\left\{\text{Ei}\left(2n\left[\frac{L_s}{2L_\infty}-j2\pi^2|f_2|\beta_2 L_s\frac{B_{\text{WDM}}}{2}\right]\right)\right\}+\pi e^{2n\frac{L_s}{2L_\infty}}\right]$$

Carrying out some possible simplifications, we finally get:

$$Q_1 = \frac{N_s}{2\pi^2|f_2|\beta_2 L_\infty}\tan^{-1}\left(2\pi^2|f_2|\beta_2 L_{\text{eff,a}} B_{\text{WDM}}\right) + \frac{1}{2\pi^2|f_2|\beta_2 L_\infty}\sum_{n=1}^{N_s-1}(N_s-n)$$

$$\left[-e^{n\frac{L_s}{L_{\text{eff,a}}}}\text{Im}\left\{\text{Ei}\left(-n\left[\frac{L_s}{L_\infty}-j2\pi^2|f_2|\beta_2 L_s B_{\text{WDM}}\right]\right)\right\}-\right.$$

$$\left.e^{-n\frac{L_s}{L_{\text{eff,a}}}}\text{Im}\left\{\text{Ei}\left(n\left[\frac{L_s}{L_\infty}-j2\pi^2|f_2|\beta_2 L_s B_{\text{WDM}}\right]\right)\right\}+\pi e^{n\frac{L_s}{L_\infty}}\right]$$

*Eq. 25*

Substituting $Q_1$ from Eq. 25 back into Eq. 8, we then get the PSD of NLI at frequency $f=0$ as:

$$G_{\text{NLI}}(0) \approx \frac{16}{27}\gamma^2 G_0^3 L_{\text{eff}}^2 N_s \int_{-B_{\text{WDM}}/2}^{B_{\text{WDM}}/2} \frac{1}{2\pi^2|f_2|\beta_2 L_{\text{eff,a}}}\tan^{-1}\left(2\pi^2|f_2|\beta_2 L_{\text{eff,a}} B_{\text{WDM}}\right)+$$

$$+\frac{16}{27}\gamma^2 G_0^3 L_{\text{eff}}^2 \int_{-B_{\text{WDM}}/2}^{B_{\text{WDM}}/2} \frac{1}{2\pi^2|f_2|\beta_2 L_{\text{eff,a}}}\sum_{n=1}^{N_s-1}(N_s-n)$$

$$\left[-e^{n\frac{L_s}{L_{\text{eff,a}}}}\text{Im}\left\{\text{Ei}\left(-n\left[\frac{L_s}{L_{\text{eff,a}}}-j2\pi^2|f_2|\beta_2 L_s B_{\text{WDM}}\right]\right)\right\}-\right.$$

$$\left.e^{-n\frac{L_s}{L_{\text{eff,a}}}}\text{Im}\left\{\text{Ei}\left(n\left[\frac{L_s}{L_{\text{eff,a}}}-j2\pi^2|f_2|\beta_2 L_s B_{\text{WDM}}\right]\right)\right\}+\pi e^{n\frac{L_s}{L_{\text{eff,a}}}}\right]df_2$$

*Eq. 26*

The first integral in Eq. 26 derives from the first term in Eq. 15 and shows a direct proportionality with respect to the number of spans $N_s$. Under the assumption of *incoherent NLI accumulation* at the end of the link (see [11], Sect. XI.C), this would be the only contribution to $G_{\text{NLI}}(0)$. It amounts to the assumption that the NLI produced in each span sums up in power at the end of the link, from which the obvious proportionality to the number of spans $N_s$. The second integral in Eq. 26 derives from the summation term in Eq. 15 and accounts for the *coherence correction* to the amount of NLI at the end of the link, i.e., it accounts for the effect of coherent interference of the NLI contributions originating at each span and beating coherently (in field and not in power) at the end of the link.

Based on the above reasoning, we rewrite Eq. 26 as the sum of the two contributions, similarly to Eq. 17 in [11]:

$$G_{\text{NLI}}(0) = G_{\text{NLI}}^{\text{inc}}(0) + G_{\text{NLI}}^{\text{cc}}(0)$$

*Eq. 27*

where the *incoherent contribution* to NLI is:

$$G_{\text{NLI}}^{\text{inc}}(0) \approx N_s \cdot \frac{16}{27} \gamma^2 G_0^3 L_{\text{eff}}^2 \int_{-B_{\text{WDM}}/2}^{B_{\text{WDM}}/2} \frac{1}{2\pi^2 |f_2| \beta_2 L_\infty} \tan^{-1}\left(2\pi^2 |f_2| \beta_2 L_\infty B_{\text{WDM}}\right) df_2$$

*Eq. 28*

and the *coherence correction contribution* is:

$$G_{\text{NLI}}^{\text{cc}}(0) \approx \frac{16}{27} \gamma^2 G_0^3 L_{\text{eff}}^2 \, 2\sum_{n=1}^{N_s-1} (N_s - n) \int_{-B_{\text{WDM}}/2}^{B_{\text{WDM}}/2} \frac{1}{4\pi^2 |f_2| \beta_2 L_\infty}$$

$$\left[ -e^{n\frac{L_s}{L_\infty}} \operatorname{Im}\left\{ \operatorname{Ei}\left( -n\left[ \frac{L_s}{L_\infty} - j2\pi^2 |f_2| \beta_2 L_s B_{\text{WDM}} \right] \right) \right\} - \right.$$

$$\left. e^{-n\frac{L_s}{L_\infty}} \operatorname{Im}\left\{ \operatorname{Ei}\left( n\left[ \frac{L_s}{L_\infty} - j2\pi^2 |f_2| \beta_2 L_s B_{\text{WDM}} \right] \right) \right\} + \pi e^{n\frac{L_s}{L_\infty}} \right] df_2$$

*Eq. 29*

The integration of Eq. 28) can be performed analytically and the exact result is also reported in [11], Appendix F, Eq. (38). It is:

$$G_{\text{NLI}}^{\text{inc}}(0) \approx N_s \cdot \frac{8}{27} \gamma^2 G_0^3 L_{\text{eff}}^2 \frac{j\left[ \operatorname{Li}_2\left(-j\pi^2 \beta_2 L_\infty B_{\text{WDM}}^2\right) - \operatorname{Li}_2\left(j\pi^2 \beta_2 L_\infty B_{\text{WDM}}^2\right) \right]}{\pi^2 \beta_2 L_\infty}$$

*Eq. 30*

We point out that the sum of the two polylogarithm functions of order 2, that is $\operatorname{Li}_2$, can be approximated with better known functions such as "log" or "asinh", to a high degree of accuracy, as either Eqs. 13 and 14 in [11].

Regarding Eq. 29, an analytical integration result could not be found, but the following quite remarkable result holds:

$$\lim_{y \to \infty} \frac{y}{2x} \cdot \left[ -e^y \operatorname{Im}\{\operatorname{Ei}(-y+jx)\} - e^{-y} \operatorname{Im}\{\operatorname{Ei}(y-jx)\} + \pi e^y \right] = \frac{\sin(x)}{x}$$

$$x, y \in \mathbb{R}^+$$

*Eq. 31*

Assuming that $y$ is "large enough", then it would be possible to use Eq. 31 as an approximation, that is:

$$\frac{y}{2x} \cdot \left[ -e^y \operatorname{Im}\{\operatorname{Ei}(-y+jx)\} - e^{-y} \operatorname{Im}\{\operatorname{Ei}(y-jx)\} + \pi e^y \right] \approx \frac{\sin(x)}{x}$$

*Eq. 32*

Notice that in [11] the above relation was given as an approximation. However, we point out that it can be proved that the exact limit Eq. 31 holds, which provides a stronger justification the approximation Eq. 32.

This approximation can be related to the integrand function of the integral in Eq. 29, by substituting:

$$x = 2\pi^2 n |f_2| \beta_2 L_s B_{\text{WDM}} \quad , \quad y = n \frac{L_s}{L_\infty}$$

*Eq. 33*

to obtain:

$$\frac{1}{4\pi^2 |f_2|\beta_2 L_\infty} \cdot \left[ -e^{n\frac{L_s}{L_\infty}} \operatorname{Im}\left\{ \operatorname{Ei}\left( -n\left[\frac{L_s}{L_\infty} - j2\pi^2 |f_2|\beta_2 L_s B_{\mathrm{WDM}}\right]\right)\right\} - \right.$$
$$\left. e^{-n\frac{L_s}{L_\infty}} \operatorname{Im}\left\{ \operatorname{Ei}\left( n\left[\frac{L_s}{L_\infty} - j2\pi^2 |f_2|\beta_2 L_s B_{\mathrm{WDM}}\right]\right)\right\} + \pi e^{n\frac{L_s}{L_\infty}} \right]$$
$$\approx \frac{\sin\left(2n\pi^2 |f_2|\beta_2 L_s B_{\mathrm{WDM}}\right)}{2n\pi^2 |f_2|\beta_2 L_s}$$

<div align="center">Eq. 34</div>

Substituting Eq. 34 into Eq. 29 we get a substantially simpler expression:

$$G_{\mathrm{NLI}}^{\mathrm{cc}}(0) \approx \frac{16}{27}\gamma^2 G_0^3 L_{\mathrm{eff}}^2 \, 2\sum_{n=1}^{N_s-1}(N_s-n)\int_{-B_{\mathrm{WDM}}/2}^{B_{\mathrm{WDM}}/2}\frac{\sin\left(2n\pi^2|f_2|\beta_2 L_s B_{\mathrm{WDM}}\right)}{2n\pi^2|f_2|\beta_2 L_s}\mathrm{d}f_2$$

<div align="center">Eq. 35</div>

Remarking now that $\sin(|x|)/|x| = \sin(x)/x$, and also noting that we can drop the subscript "2" from $f_2$, we can re-write Eq. 35 as:

$$G_{\mathrm{NLI}}^{\mathrm{cc}}(0) \approx \frac{16}{27}\gamma^2 G_0^3 L_{\mathrm{eff}}^2 \, 2\sum_{n=1}^{N_s-1}(N_s-n)\int_{-B_{\mathrm{WDM}}/2}^{B_{\mathrm{WDM}}/2}\frac{\sin\left(2n\pi^2 \beta_2 L_s B_{\mathrm{WDM}} f\right)}{2n\pi^2 \beta_2 L_s f}\mathrm{d}f$$

<div align="center">Eq. 36</div>

We have numerically investigated the approximation Eq. 34 and found it to be good in most practical cases. We leave for a subsequent study a precise identification of its envelope of validity. Notice that such validity check should actually be carried out on the overall output of Eq. 36, since it is the accuracy of Eq. 36 that we are interested in, rather than that of Eq. 34, at any specific value of its variables and parameters.

In [11] then the further assumption is made: the integration bandwidth $B_{\mathrm{WDM}}$ is large enough so that we can approximate the integral in Eq. 35 as:

$$\int_{-B_{\mathrm{WDM}}/2}^{B_{\mathrm{WDM}}/2}\frac{\sin\left(2n\pi^2|f_2|\beta_2 L_s B_{\mathrm{WDM}}\right)}{2n\pi^2|f_2|\beta_2 L_s}\mathrm{d}f_2 \approx \int_{-\infty}^{\infty}\frac{\sin\left(2n\pi^2|f_2|\beta_2 L_s B_{\mathrm{WDM}}\right)}{2n\pi^2|f_2|\beta_2 L_s}\mathrm{d}f_2 = \frac{\pi}{2n\pi^2 \beta_2 L_s}$$

<div align="center">Eq. 37</div>

The results Eqs. 22-24 in [11] follow quite directly from this assumption. In the next section, we introduce new results that depart from the above assumption.

## 3. New results on closed-form coherent correction terms

The assumption Eq. 37 may not hold, especially when considering a single-channel, whereby $B_{\mathrm{WDM}}$ is replaced by the single-channel bandwidth. In the following, to stress this, we substitute:

$$B_{\mathrm{WDM}} \to B_{\mathrm{CUT}}$$

<div align="center">Eq. 38</div>

where "CUT" stands for "channel under test". Eq. 36 can then be integrated analytically, since:

$$\int_{-b}^{b} \frac{\sin(ax)}{(ax)} dx = 2 \frac{\text{SinInt}(ab)}{a}$$

*Eq. 39*

Posing: $a = 2n\pi^2 \beta_2 L_s B_{\text{WDM}}$ and $b = B_{\text{WDM}}/2$, and $x = f$, and using Eq. 39, we get:

$$\int_{-B_{\text{WDM}}/2}^{B_{\text{WDM}}/2} \frac{\sin(2n\pi^2 \beta_2 L_s B_{\text{WDM}} f)}{2n\pi^2 \beta_2 L_s f} df = B_{\text{WDM}} \int_{-B_{\text{WDM}}/2}^{B_{\text{WDM}}/2} \frac{\sin(2n\pi^2 \beta_2 L_s B_{\text{WDM}} f)}{2n\pi^2 \beta_2 L_s B_{\text{WDM}} f} df =$$

$$= B_{\text{WDM}} \int_{-B_{\text{WDM}}/2}^{B_{\text{WDM}}/2} \frac{\sin(ax)}{ax} dx = 2B_{\text{WDM}} \frac{\text{SinInt}(ab)}{a} = 2B_{\text{WDM}} \frac{\text{SinInt}(n\pi^2 \beta_2 L_s B_{\text{WDM}}^2)}{2n\pi^2 \beta_2 L_s B_{\text{WDM}}}$$

$$= \frac{\text{SinInt}(n\pi^2 \beta_2 L_s B_{\text{WDM}}^2)}{n\pi^2 \beta_2 L_s}$$

*Eq. 40*

Using Eq. 40, we can re-write Eq. 35 as:

$$G_{\text{NLI}}^{\text{cc}}(0) \approx \frac{16}{27} \gamma^2 G_0^3 L_{\text{eff}}^2 \, 2 \sum_{n=1}^{N_s-1} (N_s - n) \frac{\text{SinInt}(n\pi^2 \beta_2 L_s B_{\text{WDM}}^2)}{n\pi^2 \beta_2 L_s}$$

$$= \frac{16}{27} \frac{\gamma^2 G_0^3 L_{\text{eff}}^2}{\pi^2 \beta_2 L_s} 2 \sum_{n=1}^{N_s-1} \frac{(N_s - n)}{n} \cdot \text{SinInt}(n\pi^2 \beta_2 L_s B_{\text{WDM}}^2)$$

*Eq. 41*

Putting the incoherent and coherent NLI contributions back together, we then have the fully closed-form approximate formula:

$$G_{\text{NLI}}(0) = G_{\text{NLI}}^{\text{inc}}(0) + G_{\text{NLI}}^{\text{cc}}(0)$$

$$\approx N_s \cdot \frac{8}{27} \gamma^2 G_0^3 L_{\text{eff}}^2 \frac{j \left[ \text{Li}_2\left(-j\pi^2 \beta_2 L_\infty B_{\text{WDM}}^2\right) - \text{Li}_2\left(j\pi^2 \beta_2 L_\infty B_{\text{WDM}}^2\right) \right]}{\pi^2 \beta_2 L_\infty}$$

$$+ \frac{16}{27} \frac{\gamma^2 G_0^3 L_{\text{eff}}^2}{\pi^2 \beta_2 L_s} 2 \sum_{n=1}^{N_s-1} \frac{(N_s - n)}{n} \cdot \text{SinInt}(n\pi^2 \beta_2 L_s B_{\text{WDM}}^2)$$

$$\approx \frac{8}{27} \frac{\gamma^2 G_0^3 L_{\text{eff}}^2}{\pi^2 \beta_2} \left\{ N_s \frac{j \left[ \text{Li}_2\left(-j\pi^2 \beta_2 L_\infty B_{\text{WDM}}^2\right) - \text{Li}_2\left(j\pi^2 \beta_2 L_\infty B_{\text{WDM}}^2\right) \right]}{L_\infty} \right.$$

$$\left. + \frac{4}{L_s} \sum_{n=1}^{N_s-1} \left[ \frac{N_s}{n} - 1 \right] \cdot \text{SinInt}(n\pi^2 \beta_2 L_s B_{\text{WDM}}^2) \right\}$$

*Eq. 42*

Eq. 42 is an original result and it improves over pre-existing GN-model results as to the closed-form modeling of NLI coherence effects.

Note that Eq. 42 can be simplified through the following approximation:

$$j \left[ \text{Li}_2(-jx) - \text{Li}_2(jx) \right] \approx \pi \, \text{asinh}\left(\frac{x}{2}\right) \approx \pi \log_e(1+x)$$

*Eq. 43*

This is typically a quite accurate approximation. So, the term $G_{\text{NLI}}^{\text{inc}}(0)$ can be recast in terms of either asinh or a log function, with typically little loss of accuracy. This was dealt with in depth in both [11] and [16] and will not be further addressed here.

## Use for single-channel NLI estimation

The closed-form formula Eq. 42 can be used for single-channel NLI estimation (or SCI, single-channel interference). It is enough to assign the single-channel bandwidth to $B_{\text{WDM}}$. Specifically, if the single-channel is the one of interest, or the "channel under test" (CUT), then: $B_{\text{WDM}} = B_{\text{CUT}}$. We omit to discuss the implications of non-zero roll-off, but point out that the impact of a non-zero roll-off is modest as long as its value is below 0.2-0.3.

We can then write:

$$G_{\text{NLI}}(0) = G_{\text{NLI}}^{\text{inc}}(0) + G_{\text{NLI}}^{\text{cc}}(0)$$

$$G_{\text{NLI}}^{\text{inc}}(0) \approx \frac{8}{27} \frac{\gamma^2 G_0^3 L_{\text{eff}}^2}{\pi^2 \beta_2} N_s \frac{j\left[\text{Li}_2\left(-j\pi^2 \beta_2 L_\infty B_{\text{CUT}}^2\right) - \text{Li}_2\left(j\pi^2 \beta_2 L_\infty B_{\text{CUT}}^2\right)\right]}{L_\infty}$$

$$G_{\text{NLI}}^{\text{cc}}(0) \approx \frac{8}{27} \frac{\gamma^2 G_0^3 L_{\text{eff}}^2}{\pi^2 \beta_2} \frac{4}{L_s} \sum_{n=1}^{N_s-1}\left[\frac{N_s}{n} - 1\right] \cdot \text{SinInt}\left(n\pi^2 \beta_2 L_s B_{\text{CUT}}^2\right)$$

*Eq. 44*

Regarding the coherent correction term, $G_{\text{NLI}}^{\text{cc}}(0)$, further simplification is possible, under further approximation. The reason for seeking a simpler form is the presence in $G_{\text{NLI}}^{\text{cc}}(0)$ of a summation that cannot be summed in closed form, but needs to be evaluated term by term. We first remark that the SinInt function can be roughly approximated, for positive values of its argument, as $\text{Si}_{\text{app}}(x)$:

$$\text{SinInt}(x) \approx \text{Si}_{\text{app}}(x) = \begin{cases} x, & 0 \leq x < \frac{\pi}{2} \\ \frac{\pi}{2}, & x \geq \frac{\pi}{2} \end{cases}$$

*Eq. 45*

We can then write:

$$G_{\text{NLI}}^{\text{cc}}(0) \approx \frac{8}{27} \frac{\gamma^2 G_0^3 L_{\text{eff}}^2}{\pi^2 \beta_2} \frac{4}{L_s} \sum_{n=1}^{N_s-1}\left[\frac{N_s}{n} - 1\right] \cdot \text{Si}_{\text{app}}\left(n\pi^2 \beta_2 L_s B_{\text{CUT}}^2\right)$$

*Eq. 46*

In many instances, the argument of $\text{Si}_{\text{app}}(x)$ will be greater than $\pi/2$, even for $n=1$ in the summation. In that case, according to the definition, $\text{Si}_{\text{app}}(x) = \pi/2$. If so, we then have:

$$G_{\text{NLI}}^{\text{cc}}(0) \approx \frac{8}{27} \frac{\gamma^2 G_0^3 L_{\text{eff}}^2}{\pi^2 \beta_2} \frac{4}{L_s} \frac{\pi}{2} \sum_{n=1}^{N_s-1}\left[\frac{N_s}{n} - 1\right]$$

*Eq. 47*

To provide an example, assuming $\beta_2 = 21$ ps²/km (typical of SMF), $L_s = 100$ km and $B_{\text{CUT}} = 32$ GBaud, then $\pi^2 \beta_2 L_s B_{\text{CUT}}^2 = 21 \gg \pi/2$. That is, $\text{Si}_{\text{app}}\left(n\pi^2 \beta_2 L_s B_{\text{CUT}}^2\right) = \pi/2$, even for $n=1$. Interestingly, the summation can be written as:

$$\sum_{n=1}^{N_s-1}\left[\frac{N_s}{n}-1\right]=1-N_s+N_s\,\mathrm{HarNum}(N_s-1)$$

*Eq. 48*

which in turn could be further simplified by resorting to: $\mathrm{HarNum}(m)\approx\log(m)+\gamma_e$, where $\gamma_e$ is Euler's constant, approximately 0.577.

If dispersion was substantially lower, as in the case of NZDSF or even DS fibers, then it is possible that $\pi^2\beta_2 L_s B_{\mathrm{CUT}}^2 < \pi/2$. In that case Eq. 47 is not a viable approximation and Eq. 46 should be used.

A way to simplify the evaluation of Eq. 46 is that of remarking that $\mathrm{Si}_{\mathrm{app}}(x)$ is a non-decreasing function of its argument. So we can write:

$$\mathrm{Si}_{\mathrm{app}}(x)\leq \mathrm{Si}_{\mathrm{app}}(n\cdot x)\quad,\quad n\in\mathbb{Z}^+$$

*Eq. 49*

As a result, we can write an *approximate lower bound of Eq. 46* by substituting the left-hand side of Eq. 49 for the right-hand side:

$$\begin{aligned}G_{\mathrm{NLI}}^{\mathrm{cc}}(0) &>\approx \frac{8}{27}\frac{\gamma^2 G_0^3 L_{\mathrm{eff}}^2}{\pi^2\beta_2}\frac{4}{L_s}\mathrm{Si}_{\mathrm{app}}\left(\pi^2\beta_2 L_s B_{\mathrm{CUT}}^2\right)\sum_{n=1}^{N_s-1}\left[\frac{N_s}{n}-1\right]\\ &=\frac{8}{27}\frac{\gamma^2 G_0^3 L_{\mathrm{eff}}^2}{\pi^2\beta_2}\frac{4}{L_s}\mathrm{Si}_{\mathrm{app}}\left(\pi^2\beta_2 L_s B_{\mathrm{CUT}}^2\right)\left[1-N_s+N_s\,\mathrm{HarNum}(N_s-1)\right]\\ &=N_s\cdot\frac{8}{27}\frac{\gamma^2 G_0^3 L_{\mathrm{eff}}^2}{\pi^2\beta_2}\frac{4}{L_s}\mathrm{Si}_{\mathrm{app}}\left(\pi^2\beta_2 L_s B_{\mathrm{CUT}}^2\right)\left[\frac{1-N_s}{N_s}+\mathrm{HarNum}(N_s-1)\right]\end{aligned}$$

*Eq. 50*

Eq. 50 does not contain any summations and represent a quite simple form of the coherence correction term, albeit approximate.

### Use of Eq. 50 in extended contexts

Eq. 50 was used to write Eq. 3 in [20] and specifically the coherence correction term, which is highlighted in [20] by a circle marker. Note that [20] deals with systems that, in general, do not have identical spans. To use the equation in that context, it was assumed that the coherence correction could be referred to each single span (say, the *n*-th) and expressed for that span in terms of the parameters specific to that span. In other words:

$$G_{\mathrm{NLI}}^{\mathrm{cc}}(0)=\sum_{n=1}^{N_s}G_{\mathrm{NLI},n}^{\mathrm{cc}}(0)$$

*Eq. 51*

where:

$$G_{\mathrm{NLI},n}^{\mathrm{cc}}(0)=\frac{8}{27}\frac{\gamma_n^2 G_{0,n}^3 L_{\mathrm{eff},n}^2}{\pi^2\beta_{2,n}}\frac{4}{L_{s,n}}\mathrm{Si}_{\mathrm{app}}\left(\pi^2\beta_{2,n} L_{s,n} B_{\mathrm{CUT}}^2\right)\left[\frac{1-N_s}{N_s}+\mathrm{HarNum}(N_s-1)\right]$$

*Eq. 52*

Clearly Eq. 51 and Eq. 52 represent coarse approximations. On the other hand, a posteriori, over the large system set addressed in [20], they appear to improve results. Note also that in Eq. 3 of [20], $\mathrm{SinInt}(x)$ was used rather than $\mathrm{Si}_{\mathrm{app}}(x)$ as in Eq. 52. Of course, $\mathrm{Si}_{\mathrm{app}}(x)\approx\mathrm{SinInt}(x)$, so it is always possible to revert to $\mathrm{SinInt}(x)$. The use of $\mathrm{Si}_{\mathrm{app}}(x)$ was helpful to derive Eq. 50 from Eq. 46, by means of Eq. 49. However, in the context of Eq. 52 one can revert to $\mathrm{SinInt}(x)$.

## 4. Acknowledgements

This work was sponsored by CISCO Photonics under an SRA agreement with Politecnico di Torino. The author would like to thank Stefano Piciaccia and Fabrizio Forghieri from CISCO Photonics, as well as Mahdi Ranjbar Zefreh and Andrea Carena, for the fruitful discussions and interactions.